\documentclass[%
 reprint,
 amsmath,amssymb,
 aps,
pre]{revtex4-2}

\usepackage{blkarray}

\usepackage{graphicx}
\usepackage{dcolumn}
\usepackage{bm}
\usepackage{hyperref}

\usepackage{xcolor}

\begin{document}


\title{$Q$-neighbor Ising model on a polarized network}

 \author{Anna Chmiel and Julian Sienkiewicz}
 \affiliation{Faculty of Physics, Warsaw University of Technology, Koszykowa 75, 00-662 Warszawa, Poland}
 \email{julian.sienkiewicz@pw.edu.pl}






\begin{abstract}
In this paper, we examine the interplay between the lobby size $q$ in the $q$-neighbor Ising model of opinion formation [Phys. Rev. E {\bf 92}, 052105] and the level of overlap $v$ of two fully connected graphs. Results suggest that for each lobby size $q \ge 3$, a specific level of overlap $v^*$ exists, which destroys initially polarized clusters of opinions. By performing Monte-Carlo simulations, backed by an analytical approach, we show that the dependence of the $v^*$ on the lobby size $q$ is far from trivial in the absence of temperature, showing consecutive maximum and minimum, that additionally depends on the parity of $q$. The temperature is, in general, a destructive factor; its increase leads to the collapse of polarized clusters for smaller values of $v$ and additionally brings a substantial decrease in the level of polarization. However, we show that this behavior is counter-intuitively inverted for specific lobby sizes and temperature ranges.  

\end{abstract}

\maketitle

\section{Introduction}
The most fascinating and at the same time well-known property of all complex systems is the impossibility to predict \textit{ad hoc} macroscopic properties of the system given the rules that govern its behavior in the micro-scale. Nonetheless, typically, based on similarities among the models we should be to verbalize at least our quantitative expectations with respect to the system in question.  

However, opinion formation models bring to that front the fact that even the duplication of the topological layer on which the dynamics takes place can lead to surprising outcomes. Vivid examples of this thesis are $q$-neighbor Ising \cite{PhysRevE.92.052105} and $q$-voter \cite{Nyczka} models that modify the original kinetic Ising model \cite{Macy2024} and voter model \cite{Clifford1973} by restricting the number of interacting neighbors. The introduction of the second level (i.e., a duplex multiplex network) changes the type of the phase transition for a given parameter \cite{Chmiel2017,chmiel}, switching from continuous to discontinuous one. In the same way, creating an asymmetry, either in the form of overlap between layers \cite{Chmiel2017} or by imposing different values of $q$ on different levels \cite{Chmiel2020} can lead to mixed-order or consecutive phase transitions. Thus, the behavior of the $q$-lobby models has lately been thoroughly examined in different settings \cite{Chmiel2018,Gradowski2020,Lou2021,Abramiuk2021,Arek2022,Krawiecki2023,Krawiecki2024}.       

Apart from strictly statistical-physics-driven motivations, this study also references social phenomena. The problem of opinion polarization is currently one of the most discussed topics in science, and a lot of effort is being made to understand its origins and threats connected to it and create possible effective countermeasures \cite{Waller2021,Voelkel2023,Combs2023}. Apart from data-driven approaches seen from the sociological point of view \cite{dimaggio1996have,mccright,mouw2001culture}, several agent-based and stochastic models have been created \cite{lambiotte2007majority,Baumann_2020,Baumann2021,Gajewski2022} to consider possible scenarios. The advent of the popular use of Artificial Intelligence (AI) and in particular, the overwhelming adoption of Large Language Models in everyday use raises questions about the role of algorithmic bias and recommendation systems leading to recent studies modeling these issues \cite{Perlata2021,Bellina2023}. The same effect can be attributed to information overload (IOL), understood as a contamination of our information space, leading in turn to masking of some of the data and, in this way, creating information bubbles (cocoons) \cite{Holyst2024,Baumann_2020}.

In this study, we consider the behavior of the $q$-neighbour Ising model in the topology setting described by Lambiotte \textit{et al.} \cite{lambiotte2007majority,lambiotte2007coexistence}. In this way, we take into account two important aspects of opinion formation models: (1) the issue of the lobby that influences our decisions (manifested by the value of $q$ the $q$-neighbour Ising model) and (2) the conditions that need to be fulfilled for polarized groups to break (manifested by the overlap of communities in the Lambiotte model). The main aim is to examine the complex interplay between these parameters and, in particular, to understand the role of the lobby size in breaking the polarization. Last but not least, we also focus on the temperature parameter in the $q$-neighbour Ising model and examine its influence on breaking or counter-intuitively sustaining the polarization of overlapping communities. We finish by considering the infinite range case and discussing the ramifications of the obtained results.

\section{Model setting}\label{sec:setting}

\begin{figure}[!hb]
         \includegraphics[width=\columnwidth]{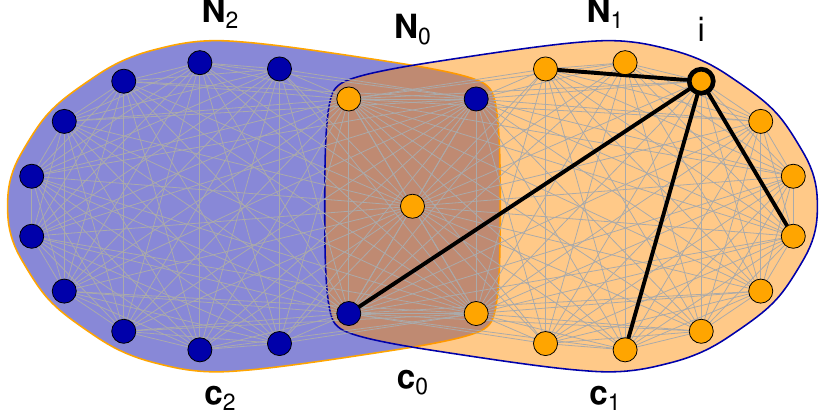}
\caption{An illustrative example of the examined system with $N_c=15$ nodes in each clique (dark-blue and orange outlines) and $v=1/3$. In such a setting $N_1 = N_2 = (1-v)N_c = 10$, while $N_0=vN_c=5$ (overlapping part) and the total number of nodes is $N=(2-v)N_c=25$. The color of the node refers to the initial state: $+1$ (orange) or $-1$ (blue). The picture presents the process of a node $i$ state update for $q=4$: a random node is selected, and out of its 15 possible neighbors (orange cluster and the interconnected one), 4 nodes are drawn at random. Depending on the total spin of the neighbors as well as on the temperature, the spin shall flip or stay in its original state (here, if $T=0$, the node cannot change its state).}
\label{fig:scheme}
\end{figure}

Lambiotte \textit{et al.} \cite{lambiotte2007majority} introduced the topology of coupled fully-connected networks or simply two overlapping cliques (here we consider a case where both cliques have the same number of nodes $N_c$). Obviously, some nodes of the first clique share connections only among themselves and the overlapping part (we will call it the first cluster; see right-hand side of Fig.~\ref{fig:scheme}). In the same manner, some nodes of the second clique share connections only among themselves and the overlapping part (we will call it the second cluster; see left-hand side of Fig.~\ref{fig:scheme}). The remaining part (the overlap, denoted further with index "0", consisting of $N_0$ nodes; see the center of Fig.~\ref{fig:scheme}) acts as an interface between these two clusters -- direct connections between cluster one and cluster two are not possible in this system. To easily parameterize the system, a control parameter $v=N_0 / N_c$ is introduced, thus the number of nodes in the first cluster is equal to $N_1 = N_2 = (1-v)N_c$ and the total number of nodes in the system expressed in terms of $N_c$ is $N = N_1 + N_2 + N_0 = (2-v)N_c$.

The idea behind the original work is to examine the behavior of the majority vote rule (MR) model with $G=3$ neighbors in this topology. However, the initial conditions are imposed in the following way: all the nodes in the first cluster have opinion $+1$, all the nodes in the second one share opinion $-1$, and the vertices in the interconnected part take $+1$ or $-1$ with equal probabilities. In this way, polarization among the first and the second cluster is maintained using the interface nodes. However, once $v$ crosses some critical value, one of the clusters ``convinces'' the other, and all the nodes in the system take the same opinion.

Unlike the original paper \cite{lambiotte2007majority}, we focus our attention on the so-called $q$-neighbor Ising model, described in detail in \cite{PhysRevE.92.052105} -- the system consists of spins $s_i = \pm 1$ that, in contrast to the original Ising model,  interact only with $q$ neighbors, selected randomly in each timestep (we called this set of neighbors a $q$-lobby). Therefore the basic dynamics of the model follows these steps: (1) randomly choose a spin $s_i$ and from all its neighbors choose a subset of $q$ neighbors, $nn_q$, (2) calculate the change of the ``energy'' related to the potential flip of spin $s_i$, i.e., $\Delta E = E(-s_i)-E(s_i)=2s_i\sum_{j \in nn_i}s_j$, (3) flip the spin with probability $\min[1,e^{-\Delta E/T}]$. The principal idea behind such a modification of the original Ising model is that, in reality, an individual cannot interact with all their neighbors in the system.  

In the following calculations, we tacitly assume that $N \rightarrow \infty$, i.e., we focus on the parameter $c$ reflecting the concentration of ``up'' spins. Concentration can be easily transformed into magnetization as $m=2c-1$; $c_1$ shall denote concentration in the first cluster, $c_2$ -- in the second one, and $c_0$ -- in the interconnected part (see notation in Fig.~\ref{fig:scheme}). Let us first write equations for the change of the concentration $c_1$ of spins in the first cluster. The central node has to belong to the first clique (see the node marked as $i$ in Fig.~\ref{fig:scheme}). At the same time, its $q$ neighbors can be selected with probability $1-v$ from the first clique and probability $v$ from the interconnecting nodes, characterized by spin concentration equal to $c_0$ (formally, the probabilities are $v/(v+1-v)$ and $(1-v)/(v+1-v)$, but as $v+1-v=1$ they can be simplified). Thus the transition probability $\gamma_1^{+}$ that number of ``up`` spins increases by one, and transition probability $\gamma_1^{-}$, that the number of ``up'' spins is decreased by one are, respectively,


\begin{widetext}
\begin{equation}
\left\{
\begin{aligned}
\gamma_1^{+} &= (1-c_1) \sum\limits_{j=0}^{q} {q \choose j} (1-v)^{q-j}v^j \sum_{k=0}^{q-j}{q-j \choose k}c_1^{q-j-k}(1-c_1)^k e_{k,q} 
\sum_{k'=0}^j{j \choose k'}c_0^{j-k'}(1-c_0)^{k'} e_{k',q}\\
\gamma_1^{-} &= c_1 \sum\limits_{j=0}^{q} {q \choose j} (1-v)^{q-j}v^j
\sum_{k=0}^{q-j}{q-j \choose k}(1-c_1)^{q-j-k}c_1^k e_{k,q}  
\sum_{k'=0}^j{j \choose k'}(1-c_0)^{j-k'}c_0^{k'} e_{k',q}    
\end{aligned}\right.
\end{equation}
\end{widetext}

where $e_{k,q} e_{k',q} = E_{k+k',q} = \min \left\{1, \exp\left[ \frac{2q-4(k+k')}{T}\right] \right\}$. Due to the symmetry of the problem, equations for the change of concentration $c_2$ are the same, except for the simple fact that one needs to swap $c_1$ with $c_2$:

\begin{widetext}
\begin{equation}
\left\{
\begin{aligned}
\gamma_2^{+} &= (1-c_2) \sum\limits_{j=0}^{q} {q \choose j} (1-v)^{q-j}v^j \sum_{k=0}^{q-j}{q-j \choose k}c_2^{q-j-k}(1-c_1)^k e_{k,q} 
\sum_{k'=0}^j{j \choose k'}c_0^{j-k'}(1-c_0)^{k'} e_{k',q}\\
\gamma_2^{-} &= c_2 \sum\limits_{j=0}^{q} {q \choose j} (1-v)^{q-j}v^j
\sum_{k=0}^{q-j}{q-j \choose k}(1-c_2)^{q-j-k}c_1^k e_{k,q}  
\sum_{k'=0}^j{j \choose k'}(1-c_0)^{j-k'}c_0^{k'} e_{k',q}    
\end{aligned}\right.
\end{equation}
\end{widetext}


When describing the changes in concentration in the common (interconnected) cluster we follow a similar path. However, this time, the central node needs to be in the interconnected cluster, and its neighbors are situated among other nodes (thus, belonging to the first cluster, the second, the interconnected, or any combination of these three). Thus, unlike the previous case, we have the probability $v/(2-v)$ for the neighbor to be in the interconnected cluster and $(1-v)/(2-v)$ to be in any of the two remaining groups. This leads to much more complicated expressions of concentration changes:  

\begin{widetext}
\begin{equation}
\left\{
\begin{aligned}
\begin{split}
\gamma_0^{+} &= (1-c_0) \sum\limits_{j=0}^{q} {q \choose j} \left(\frac{1-v}{2-v} \right)^{q-j}\left(\frac{v}{2-v}\right)^j 
\sum_{k=0}^{q-j}{q-j \choose k}c_0^{q-j-k}(1-c_0)^k e_{k,q}\times\\ 
& \sum_{l=0}^{j}{j \choose l}\left[ \sum_{m=0}^{j-l}{j-l \choose m}c_1^{j-l-m}(1-c_1)^m e_{m,q}\sum_{m'=0}^j{l \choose m'}c_2^{l-m'}(1-c_2)^{m'} e_{m',q}\right]\\
\gamma_0^{-} &= c_0 \sum\limits_{j=0}^{q} {q \choose j} \left(\frac{1-v}{2-v} \right)^{q-j}\left(\frac{v}{2-v}\right)^j \sum_{k=0}^{q-j}{q-j \choose k}(1-c_0)^{q-j-k}c_0^k e_{k,q}\times\\ 
&\sum_{l=0}^{j}{j \choose l}\left[ \sum_{m=0}^{j-l}{j-l \choose m}(1-c_1)^{j-l-m}c_1^m e_{m,q}\sum_{m'=0}^j{l \choose m'}(1-c_2)^{l-m'}c_2^{m'} e_{m',q}\right]
\end{split}
\end{aligned}\right.
\end{equation}
\end{widetext}

Following, we shall provide results of the analytical approach backed by numerical simulations of the model. For the sake of simplicity, we shall first consider the example of vanishing temperature (i.e., $T \rightarrow$ 0). After examining the behavior for different values of $q$, we will move to a more general case with $T > 0$.

\section{$T \rightarrow 0$ case}
In order to find the stationary solution for $c_1$ we need to solve the following equation:
\begin{equation}
    \langle c_1(t+1) \rangle - \langle c_1(t) \rangle = \gamma_1^{+} - \gamma_1^{-} = 0.
\label{eq:c1}
\end{equation}
In general, the solution leads to an equation combining $c_1$ and $c_0$; however, following \cite{lambiotte2007majority} and the results obtained from simulations, we shall set $c_0=1/2$ and rewrite $c_1$ as $1/2 + x$. In the case of a low number of neighbours $q$, it is possible to obtain relatively simple forms of Eq. (\ref{eq:c1}) and thus closed solutions. In particular, for $q=1$ and $q=2$, the system does not have any stationary solution apart from the paramagnetic state (see Appendix \ref{app:app1}).

\subsection{Exact solutions for q=3,4,5}
The first non-trivial case can be obtained when $q=3$. In this situation Eq.~(\ref{eq:c1}), 
after applying the transformation $c_0 \rightarrow 1/2$ and  $c_1 \rightarrow x + 1/2$, takes the form of a cubic equation that can be factorized into 
\begin{equation}
    x(A x^2 + B) = 0.
\label{eq:cubic}
\end{equation}
with an obvious set of solutions $x_0 = 0$ and $x_{1,2} = \pm \sqrt{-B/A}$, where $A = 4(1-v)^2$ and $B = 1-3v$, leading to

\begin{equation}
\left\{
\begin{aligned}
x_0 ~ ~ &= ~ 0\\
x_{1,2} &= \pm \frac{1}{2(1-v)} \sqrt{\frac{1-3v}{1-v}}
\label{eq:q3sol}
\end{aligned}\right.
\end{equation}

\begin{figure*}[!ht]
         \includegraphics[width=\textwidth]{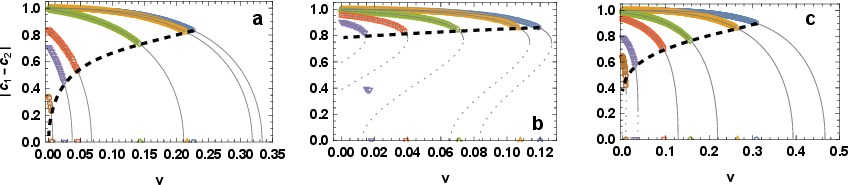}
\caption{Absolute net concentration $|c_1-c_2|$ versus $v$ for the frustrated system with (a) $q=3$, (b) $q=4$, and (c) $q=5$. Points are Monte-Carlo simulations ($N=1~000~000$, $M=2000$ large MC steps, i.e., for each $v$ we perform $2\cdot10^9$ updates of the system). (a) Solid lines are $2x_0$ and $2x_1$ given by Eq.~(\ref{eq:q3sol}), while the dashed vertical line is the critical value $v_3^*$ from Eq.~(\ref{eq:v3}). (b) Solid and dotted lines come from Eq.~(\ref{eq:q4sol}), while the dashed vertical line is the critical value $v_4^*$ from Eq.~(\ref{eq:v4}). (c) Solid and dotted lines come from Eq.~(\ref{eq:q5sol}), while the dashed vertical line showing $v_5^*$ is Eq.~(\ref{eq:v5})}
\label{fig:fig2}
\end{figure*}

The above result might suggest that the magnetization undergoes a continuous phase transition with critical values of $v$ equal to $v^*=1/3$ for which the system moves to the paramagnetic state. However, we initially assumed that the concentration in the interconnected cluster stays at $c_0=1/2$, which does not hold for an arbitrary value of $v>0$. In fact, as described in \cite{lambiotte2007majority}, the interconnected system is metastable; thus, we need to perform linear stability analysis (LSA) for the obtained solutions leading to a matrix  
\begin{equation}
    \mathbf{L} = ~\begin{blockarray}{(ccc)}
s & t & t\\
u & w & 0\\
u & 0 & w\\
\end{blockarray}
\end{equation}
containing coefficients of the linear expansion. As the last step, we need to calculate the largest eigenvalue of $L$ and determine the point where it crosses zero (see Appendix \ref{app:lsa} for details of the procedure). 



In the case of $q=3$, the coefficients of the linear expansion have a relatively simple form
\begin{equation}
\left\{
\begin{aligned}
s &= \frac{4-5v}{2(v-2)}\\
t &= \frac{3(v-1)}{2(v-2)}\\
u &= \frac{-3v^2}{v-1}\\
w &= 3v-1
\label{eq:L3}
\end{aligned}\right.
\end{equation}
and eigenvalues of $\mathbf{L}$ are then
\begin{equation}
\left\{
\begin{aligned}\lambda_0 ~ ~ &= ~ 3v-1\\
\lambda_{1,2} &= \frac{(2v-1)(3v-8) \pm 3v\sqrt{4v^2-28v+41}}{4(v-2)}
\end{aligned}\right.
\label{eq:lambdamat}
\end{equation}

with $\lambda_2$ being the largest eigenvalue for any $v \in [0, \frac{1}{3}]$. Finally, by solving $\lambda_2 = 0$ we obtain the critical point 

\begin{equation}
v_3^{*} = \frac{1}{6}\left(\sqrt{337}-17\right) \approx 0.22626  
\label{eq:v3}
\end{equation}

for which $x_1$ loses stability. Figure \ref{fig:fig2}a presents a comparison between analytical predictions and Monte-Carlo simulations for a system of $N=10^6$ nodes. All simulations are performed in the following way: we set $N$ and a specific value of $v$ that determines the size of the first, second, and interconnected clusters. Then, we set all nodes in the first cluster to be $s_i=+1$, all the nodes in the second one to be $s_i=-1$, and all the nodes in the interconnected cluster get a random spin. We run the dynamics for $M$ large MC steps (i.e., there in total $M N$ updates) and record the value of $c_1$, $c_2$, and $c_0$ by calculating averages in relevant clusters.  To overcome discrepancies arising from fluctuations, instead of $c_1$ we plot $|c_1-c_2|$ as an order parameter -- once the critical value predicted by Eq.~(\ref{eq:v3}) is reached, all the nodes, regardless of the cluster they belong to, acquire (on average) the same spin direction. The comparison shows perfect agreement between analytical predictions and simulated systems.

Following the above-described procedure, it is also possible to obtain results for the limiting case of $T \rightarrow 0$ and for $q=4$. Here, we arrive at a quintic equation that can be factorized as
\begin{equation}
    x(Ax^4 + Bx^2+C)=0
\label{eq:quintic}
\end{equation}
characterized by the following five solutions:
\begin{equation}
\left\{
\begin{aligned}
x_0 ~ ~ &= ~ 0\\
x_{1,2} &= \pm \sqrt{-\frac{1}{2A}\left(B - \sqrt{B^2-4 A C} \right)}\\
x_{3,4} &= \pm \sqrt{-\frac{1}{2A}\left(B + \sqrt{B^2-4 A C} \right)}
\end{aligned}\right.
\label{eq:quinticsol}   
\end{equation}
where $A = -48(v-1)^4$, $B=8(v-1)^2(2v+1)$ and $C=1-12v$. After some simplifications, we arrive at the following set of solutions
\begin{equation}
\left\{
\begin{aligned}
x_0 ~ ~ =& ~ 0\\
x_{1,2} =& \pm \frac{\sqrt{2 v+1 - 2\alpha_4}}{2 \sqrt{3} (1-v)}\\
x_{3,4} =& \pm \frac{\sqrt{2 v+1 + 2\alpha_4}}{2 \sqrt{3} (1-v)}
\end{aligned}\right.
\label{eq:q4sol}
\end{equation}
where $\alpha_4 = \sqrt{v(v-8) + 1}$. The above set of equations is a typical example of a system with hysteresis: as long as $v < 1/12$ there are only three real solutions ($x_0$, $x_3$ and $x_4$) in the range of $v \in [1/12, 4-\sqrt{15}]$ all five solutions are real, and once $v > 4-\sqrt{15}$ only $x = 0$ exists (see solid lines in Fig.~\ref{fig:fig2}b). However, once again, to find the critical value of $v$, one needs to follow linear stability analysis. In this case, the coefficients of matrix $\mathbf{L}$ are given by
\begin{equation}
\left\{
\begin{aligned}
s &= \frac{22-23v}{8(v-2)}\\
t &= \frac{3(v-1)}{2(v-2)}\\
u &= \frac{v}{3(1-v)}\left[ (4-v)(1-v)-\alpha_v(v+2)-2\alpha_v^2\right]\\
w &= -\frac{\alpha_v}{3}\left(1+2\alpha_v+2v\right)
\label{eq:Lcoeff4}
\end{aligned}\right.
\end{equation}
The eigenvalues of the matrix $\mathbf{L}$ are much more complicated than in the $q=3$ case, however, it is still possible to obtain an approximate value for $v^{*}$, namely
\begin{equation}
v_4^{*} \approx 0.1195  
\label{eq:v4}
\end{equation}
by solving $2ut-ws=0$ numerically. Figure \ref{fig:fig2}b shows a comparison of the numerical simulations and analytical solutions given by Eq.~(\ref{eq:q4sol}) and Eq.~(\ref{eq:v4}), presenting a perfect agreement of these two approaches again.

Interestingly, also the case of $q=5$, the system can be described by a quintic equation (\ref{eq:quintic}) with $A = -48(v-1)^5$, $B=40(v-1)^3$ and $C=7-15v$. By using the substitution $z=(v-1)x$ one immediately arrives at the following set of solutions    
\begin{equation}
\left\{
\begin{aligned}
x_0 ~ ~ =& ~ 0\nonumber\\
x_{1,2} =& \pm \frac{1}{2(1-v)}\sqrt{\frac{5}{3} + \frac{2}{3}\frac{\alpha_5}{1-v}}\nonumber\\
x_{3,4} =& \pm \frac{1}{2(1-v)}\sqrt{\frac{5}{3} - \frac{2}{3}\frac{\alpha_5}{1-v}}
\end{aligned}\right.
\label{eq:q5sol}
\end{equation}
where $\alpha_5 = \sqrt{(1-v)(1+5v)}$. Interestingly, the pair $x_{1,2}$ brings results outside the range $[-1/2, 1/2]$, thus only $x_0$ and $x_{3,4}$ are valid solutions for this problem. For $q=5$ coefficients of the $\mathbf{L}$ matrix read
\begin{equation}
\left\{
\begin{aligned}
s &= \frac{16-23v}{8(v-2)}\\
t &= \frac{15(v-1)}{8(v-2)}\\
u &= \frac{5v}{3(1-v)}\left( \alpha_5-1-2v\right)\\
w &= -\frac{\alpha_v}{3}\left(2+10v-5\alpha_5\right)
\label{eq:Lcoeff5}
\end{aligned}\right.
\end{equation}
Once again, by numerically solving the equation connected to the largest eigenvalue, i.e., $2ut-sw=0$ we can arrive at the critical value
\begin{equation}
v_5^{*} \approx 0.309748,  
\label{eq:v5}
\end{equation}
fully supported by numerical simulations in Fig. \ref{fig:fig2}c.

\begin{figure}[ht]
         \includegraphics[width=0.95\columnwidth]{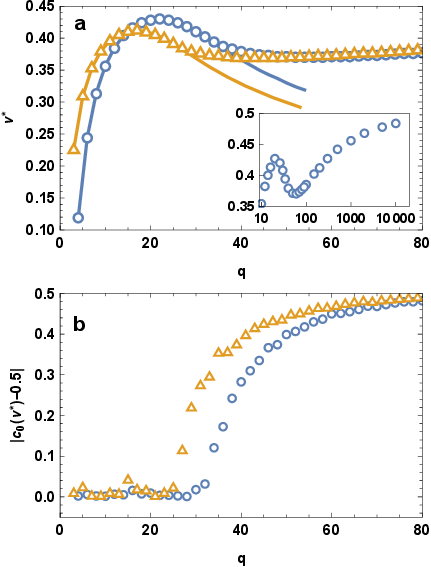}
\caption{(a) Critical value of $v$ versus the number of neighbors $q$. Data points are Monte-Carlo simulations ($N=1~000~000$, $M=2000$ large MC steps). Triangles reflect odd values of $q$ (starting from $q=3$) while circles -- even ones (starting from $q=4$). Solid lines reflect $v^*$ obtained by linear stability analysis. The inset shows a log-linear plot of $v^*(q)$ to illustrate the behavior for larger values of $q$. Due to computational constraints in this case (slow simulation time for large $q$), we used $N=100~000$ nodes and $M=1000$ large MC steps.  (b) The absolute value of shifted concentration $c_0$ in the overlapping cluster at the critical values of $v$ versus the lobby size $q$.}
\label{fig:fig3}
\end{figure}

\subsection{Higher values of $q$}
The complexity of $\gamma_1^{+} - \gamma_1^{-}$ increases with the value of $q$. Although it is still possible to write down appropriate equations, the maximum degree of the relevant polynomial is equal to $q$ for odd values of $q$ and $q+1$ for even ones. Relevant calculations, in particular the estimation of the largest eigenvalue of $\mathbf{L}$ matrix, become troublesome and connected to larger rounding errors. Therefore, to follow the system's behavior for larger values of $q$, it is easier to rely on numerical calculations than on their analytical counterpart. Figure \ref{fig:fig3}a reveals an unexpected behavior of the critical value $v^*$ when plotted against the lobby size $q$. The oscillations between odd and even values of $q$ have already been observed in the $q$-neighbor Ising model, both in the original paper \cite{PhysRevE.92.052105} as well as in \cite{Chmiel2017}. However, in this case, we encounter a non-monotonic behavior of the critical value of $v$ versus $q$, i.e., we observe clear maxima for the odd and even values correctly predicted by numerically performing relevant linear stability analysis. Additionally, although initially systems with odd $q$ were characterized with larger $v^*$ than those with $q+1$ (e.g., for $q=3$, the system stays in a polarized state for larger values of $v$ than it is the case for $q=4$), after reaching $q=15$ this behavior is inverted. As $q$ crosses 25 in the case of odd values and 35 for even ones, the agreement between numerical simulations and theoretical predictions of linear stability analysis diverges: the latter decreases with $q$ (up to the point when computations were feasible) while the former reaches a minimum and then increases toward $v^*=1/2$ (see inset in Fig. \ref{fig:fig3}a and Sec. \ref{sec:mf} for further discussion). The discrepancy between numerical simulations and the theoretical approach has its roots in the fact that when the linear stability analysis is performed, we expand the concentration in the interconnected cluster $c_0$ around $1/2$, as shown in Eq. (\ref{eq:trans}). However, Fig. \ref{fig:fig3}b clearly shows that this assumption is valid only for small values of $q$ ($q < 25$ for odd and $q < 35$ for even ones), coinciding with the region where the discrepancies first appear. Finally, let us note that for sufficiently large values of $q$, the difference between odd and even values disappears.

\begin{figure*}[ht]
\includegraphics[width=.95\textwidth]{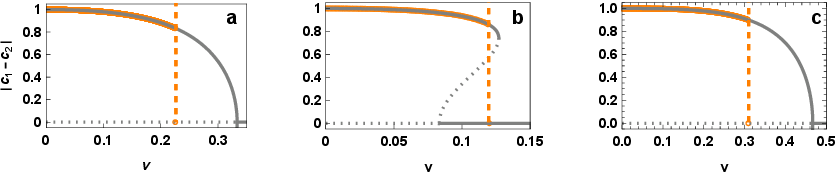}
\caption{Absolute net concentration $|c_1-c_2|$ versus $v$ for different values of temperature $T$. Points are Monte-Carlo simulations ($N=1~000~000$, $M=2000$). (a) $q=3$, right to left: $T=0$, $T=0.5$, $T=1.0$, $T=1.5$, $T=1.6$ and $T=1.7$).  Solid lines are $2x_1$ given by Eq.~(\ref{eq:3fullT}), while the dashed line comes from solving Eq.~(\ref{eq:v3Tcubic}) and substituting the obtained value into Eq.~(\ref{eq:3fullT}). (b) $q=4$, right to left $T=0$, $T=1$, $T=1.5$, $T=1.8$ and $T=2$. Solid and dotted lines come from Eq.~(\ref{eq:quinticsol}) where the coefficients are given by Eq.~(\ref{eq:coeff4}). The dashed line is obtained by substituting $v_4^*(T)$ into Eq.~(\ref{eq:quinticsol}). (b) $q=5$, right to left $T=0$, $T=1$, $T=2$, $T=2.5$, $T=3$ and $T=3.15$. Solid and dotted lines come from Eq.~(\ref{eq:quinticsol}) where the coefficients are given by Eq.~(\ref{eq:coeff5}). The dashed line is obtained by substituting $v_5^*(T)$ into Eq.~(\ref{eq:quinticsol}).}
\label{fig:fig4}
\end{figure*}

\section{The $T > 0$ case}
Until now we have deliberately omitted a crucial parameter of the $q$-neighbour Ising model -- temperature $T$. In the case of the original model \cite{PhysRevE.92.052105}, the temperature is responsible for destabilizing the system; for $q \ge 3$, the network exhibits a phase transition between ferromagnetic and paramagnetic phases at a critical temperature $T^{*}$, which linearly increases with $q$. In our case it should be safe to hypothesize that the thermal noise should act in a similar way -- the increase of $T$ ought to lower the value $v$ for which the system is still stable in its polarized state.

To check our assumptions, we shall come back to Eq.~(\ref{eq:c1}) calculated for $q=3$ but this time allowing for any $T > 0$. Despite its apparent complex form, we arrive at a cubic equation (\ref{eq:cubic}), however, in this case, its coefficients are dependent on the temperature (to simplify the notation, we further use $z=\mathrm{e}^{-\frac{2}{T}}$)

\begin{equation}
\left\{
\begin{aligned}
    A &= \left[(v-4)z^3-3(v-2)z-2(1-v)\right](1-v)^2\\
    B &= \frac{1}{4}\left[(3v-4)z^3+3(v-2)z+2(1-3v)\right]
\end{aligned}\right.
\end{equation}
and its solutions are $x_0 = 0$ as well as


\begin{equation}
x_{1,2} = \frac{\pm 1}{2(1-v)} \sqrt{\frac{(3v-4)z^3+3(v-2)z + 2(1-3v)}{(4-v)z^3+3(v-2)z + 2(1-v)}}
\label{eq:3fullT}
\end{equation}

Obviously, if $T \rightarrow 0$, i.e., $u \rightarrow 0$, we recreate, as expected, Eq. (\ref{eq:q3sol}). At this point, we can follow the procedure introduced in Appendix \ref{app:lsa} for the $T \rightarrow 0$ to obtain the critical value $v_3^*$ as a function of the temperature $v_3^*(T)$. Furthermore, by plugging $v_3^*(T)$ into Eq.~(\ref{eq:3fullT}), we are also able to calculate the predicted concentration right before the collapse of polarized clusters. Due to the complex form of the resulting equations, we refrain from presenting these solutions in an explicit form here, leaving details for Appendix~\ref{app:app3}. Figure~\ref{fig:fig4}a presents a comparison of these predictions with the data obtained directly from Monte-Carlo simulations for different values of $T$. As expected, an increase in $T$ results in a decrease in the critical value of $v$ needed to destroy polarized clusters. Figure~\ref{fig:fig4}a suggests that at a specific $T=T_c$ the value of $v_3^*$ drops down to 0. Indeed, let us note that if $v=0$ we can find $T_c$ by simply solving $-2z^3-3z+1=0$ (nominator in Eq.~(\ref{eq:3fullT}))  and inverting the introduced substitution $z=\mathrm{e}^{-\frac{2}{T}}$ to get $T_c=-2\ln^{-1}(r/2-1/r)$ where $r=[2(1+\sqrt{3})]^{1/3}$, i.e., $T_c \approx 1.7214$. This is an expected and obvious result, as in the case of $v=0$, the system is identical to the one examined in \cite{PhysRevE.92.052105}, i.e., with the lack of interface nodes, the network simply breaks down into two separate cliques with $N/2$ nodes each; thus, the model's dynamics needs to follow directly the one described in \cite{PhysRevE.92.052105}.    

Similar calculations can be performed for $q=4$ and $q=5$: in both cases we make use of Eq.~(\ref{eq:quintic}) with suitable values $A$, $B$ and $C$, denoted here with the lower index:
\begin{equation}
\left\{
\begin{aligned}
A_4 &= -16 \left(z^4-4 z^2+3\right) (v-1)^4\\
B_4 &= 8\left[-z^4 (5-2 v)+4 z^2 (1-v)+2 v+1\right](v-1)^2\\
C_4 &=z^4 (4 v-5)+4 z^2 (2 v-3)+1-12v
\end{aligned}\right.
\label{eq:coeff4}
\end{equation}
and 
\begin{equation}
\left\{
\begin{aligned}
\begin{split}A_5 &= 8\left[z^5 (v-6)-5 z^3 (v-4)+\right.\\
& \left.+ 10 z (v-2)-6 (v-1)\right](1-v)^4\\
    B_5 &= 40 \left[z^5 (v-2)-z^3 v-2 z (v-2)+2 (v-1)\right](v-1)^2\\
    C_5 &= z^5 (5 v-6)+5 z^3 (3 v-4)+10 z (v-2)-2(15v+7)
\end{split}
\end{aligned}\right.
\label{eq:coeff5}
\end{equation}
The solutions can then be found by substituting relevant coefficients in Eq.~(\ref{eq:quinticsol}). We skip presenting their explicit form as well as the details of the following procedure that leads to obtaining relations $v_4^*(T)$ and $v_5^*(T)$ due to their complex forms, however the approach is identical to the one performed for $q=3$. The results are presented in Fig.~\ref{fig:fig4}b and Fig.~\ref{fig:fig4}c. One can easily notice the profound impact of the lobby size on the behavior of $v^*(T)$. While for $q=3$, we observed negligible values of $c$ for small $v$, for $q=4$, the concentration stays almost at the constant level regardless of the thermal noise introduced into the system. For $q=5$, the decrease is more prominent, although also, in this case, the concentration does not vanish. The observations with respect to concentration at $v=0$ are expected ones -- as we already remarked, in such a situation, we are dealing with the system examined in \cite{PhysRevE.92.052105} that reports a discontinuous phase transition between the ferromagnetic and paramagnetic phase at any $q>3$. Let us note that the results reported in \cite{PhysRevE.92.052105} indicate a linear growth of the critical temperature combined with the decrease of the concentration level at the transition. Combining this observation with the results shown in the inset of Fig.~\ref{fig:fig3} allows us to expect that for large $q$ the space of $q$ should span between $0$ and $1/2$ (see also Sec. \ref{sec:mf} for further details).      

\begin{figure}[!ht]
         \includegraphics[width=\columnwidth]{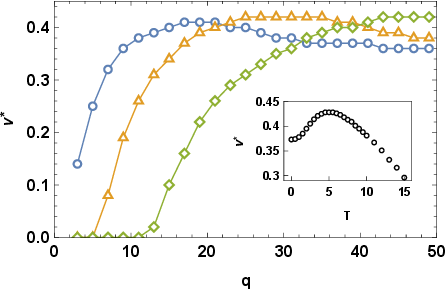}
\caption{Critical value $v^*$ versus the lobby size $q$ for different values if $T$: $T=1$ (circles), $T=4$ (triangles) and $T=10$ (diamonds). Solid lines are for eye guidance. The inset presents critical value $v^*$ versus $T$ for $q=33$ in a limited range of $T$. All data points are Monte-Carlo simulations ($N=100~000$, $M=2000$ large MC steps).}
\label{fig:fig5}
\end{figure}

Results presented in Fig.~\ref{fig:fig4} might suggest that the critical value of $v$ decreases monotonically with $T$. This is, however, not the case as proven by Fig.~\ref{fig:fig5} where we show the analog of Fig.~\ref{fig:fig3}a for different values of $T$ (for clarity of the comparison, we restrict ourselves to even values of $q$). With increasing $T$, the maximum observed in Fig.~\ref{fig:fig3} for $T=0$ moves higher lobby sizes $q$. For small values of $q$ this behavior does not bring any effect, on the other hand, it has a profound impact for $q > 21$ as depicted in the inset in Fig.~\ref{fig:fig5}, where the relation between $v^*$ and $T$ is shown for $q=33$. In this case, the initial increase of $T$ results in an increase of the critical overlap that still allows the polarization to be sustained, which is a counter-intuitive phenomenon. 

\section{The limiting case $q = N$}\label{sec:mf}

Let us finally consider the limiting case in which the number of neighbors $q$ is equal to all the nodes in the given clique, i.e., $N_1 + N_0$ in the first and second cluster and $N_1 + N_2 + N_0$ in the case of the overlapping cluster.  First, let us provide a simple argument as to why, for $T \rightarrow 0$, the critical value of $v$ should be equal to $1/2$ as conjectured from the inset in Fig. \ref{fig:fig3}a. Indeed, if $v=1/2$, then in line with Sec. \ref{sec:setting} we have $N_1 = N_2 = N_0 = N/3$. Since we prohibit any spin changes due to thermal fluctuations, as long as $N_1 > N_0$ (and $N_2 > N_0$) it is impossible to flip any spin in the first and second cluster as the sum of the spins of all the neighbors of any node in the first or second cluster is always of the same sign as the spin of the considered node. The situation is changed when $N_0 > N_1$ (and $N_0 > N_2$), thus $v=1/2$ marks the critical value.

Given the described conditions, rate equations can be written in the following way:

\begin{equation}
\left\{
\begin{aligned}
&(1-c_1) - c_1 \mathrm{e}^{\frac{-2}{~T}\left[N_1(2c_1-1)+N_0(2c_0-1)\right]} = ~ 0\\
&(1-c_2)\mathrm{e}^{\frac{-2}{~T}\left[N_1(2c_2-1)+N_0(2c_0-1)\right]} - c_2= ~ 0\\
&(1-c_0) - c_0 \mathrm{e}^{\frac{-2}{~T}\left[N_1(2c_1+2c_2-2)+N_0(2c_0-1)\right]} = ~ 0
\end{aligned}\right.
\label{eq:mf}
\end{equation}

Here, by writing out Boltzmann factors for the first and the second cliques, we assume that $c_1 > 1/2$ and $c_2 < 1/2$. We also use the same approach for the interconnected cluster, assuming that the rate for $c_0 < 1/2$ is $1$ while for $c_0 > 1/2$ it follows the relevant Boltzmann factor.

\begin{figure}[!ht]
         \includegraphics[width=\columnwidth]{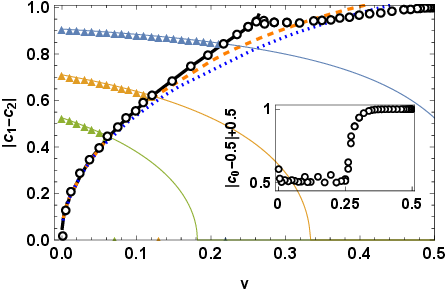}
\caption{Absolute net concentration versus $v$ in the infinite range model with $N=1~000~000$. Triangles (simulations) and thin solid lines (numerical solutions of Eq. (\ref{eq:cmf}) mark $T=300~000$, $T=400~000$ and $T=450~000$ (from top to bottom). Circles are critical values of $v$ obtained from numerical simulations for $T$ ranging between $0$ and $499~000$. Solid thick, dashed, and dotted lines are analytical predictions of $v^*$ performed using the approach described in Appendix~\ref{app:app4}, using numerical solutions of Eq. (\ref{eq:cmf}) -- solid line), Eq. (\ref{eq:appc3}) --- dashed line, and Eq. (\ref{eq:appc5}) -- dotted line. The inset presents the absolute value of shifted concentration $c_0$ in the overlapping cluster at the critical values of $v$.}
\label{fig:fig6}
\end{figure}

After using $N_1=N(1-v)/(2-v)$ and transforming the first equation together with the assumption that $c_0=1/2$ we arrive at a self-consistent equation for the concentration in the first cluster
\begin{equation}
    2c_1 - 1 = \tanh\left[\frac{N}{T}\frac{1-v}{2-v}(2c_1 - 1)\right],
\label{eq:cmf}
\end{equation}
which is equivalent to the outcome of a mean-field approach, i.e., $m=\tanh(z m/T)$, where the number of neighbors $z=N_c\frac{1-v}{2-v}$ is the total number of nodes in the first cluster. It is an expected result since the applied setting means that we consider all the nodes in the system as neighbors and treat each of them as if they were equipped with the same spin. 

Indeed, as seen in Fig.~\ref{fig:fig6} for different values of $T$, the concentration follows closely non-zero numerical solutions of the transcendental Eq.~(\ref{eq:cmf}) up the point when for some specific $v$ a transition occurs in the same way, is the previously examined cases in Sec. III and IV. The critical value (shown in empty symbols in Fig.~\ref{fig:fig6}) starts from $v^*=1/2$ for $T=0$, in line with the argument given at the beginning of this section, and goes down to $v^*=0$ for $T=T_c=N/2$ as expected from Eq.~(\ref{eq:cmf}). One has to note, though, that two distinct regions are visible in Fig. ~\ref{fig:fig6}: one between $v=0$ and $v=1/4$ and the second between $v=1/4$ and $v=1/2$. This effect is connected to the value of the concentration in the interconnected cluster $c_0$, as shown in the inset of Fig. ~\ref{fig:fig6} in the former region, it stays at $c_0=1/2$ while in the latter, it is $c_0=1$ or $c_0=0$.

To obtain theoretical predictions of $v^*(T)$, we need to perform once again the linear stability analysis described in Appendix \ref{app:lsa}. However, one cannot use it straightforwardly due to an implicit form of equations (\ref{eq:mf}). Instead, one needs to rearrange equations (\ref{eq:mf}) and expand the appearing term $\ln c - \ln(1-c)$ around specific values of $c$ (see Appendix~\ref{app:app4} for details). Despite the introduced approximations, we obtain a very good match between the numerical simulations and theoretical predictions, as can be spotted in Fig.~\ref{fig:fig6} presenting LSA results obtained by making use of numerical solutions of the transcendental Eq.~(\ref{eq:cmf}) as well as directly utilizing approximations given by Eqs.~(\ref{eq:appc3}) and (\ref{eq:appc5}).

\section{Conclusions and Discussion}
In this paper, we have examined the interplay between the lobby size $q$ in the $q$-neighbor Ising model and the level of overlap $v$ of two fully connected graphs. Numerical simulation results backed by an analytical approach indicate that for each lobby size $q \ge 3$, a specific level of overlap $v^*$ exists, which destroys initially polarized clusters of opinions. However, the dependence of the $v^*$ on the lobby size $q$ is far from trivial, showing a sequence of a maximum and a minimum before monotonically increasing toward $v=1/2$ -- this behavior additionally depends on the parity of $q$. The temperature $T$ -- a second parameter of the $q$-neighbor Ising model -- is, in general, a destructive factor for its increase leads to the earlier collapse of polarized clusters, but it additionally brings a substantial decrease in the polarization. However, counter-intuitively, for certain values of $q$ an opposite behavior is observed. Moreover, the actual value of the lobby size determines the character of the relationship between the temperature and the level of polarization.

Let us now discuss the similarities and differences between the obtained results and the original model \cite{PhysRevE.92.052105} as well as the partially overlapped duplex case \cite{Chmiel2017}. The key property of the original $q$-neighbor Ising model is the change of phase transition character: for $q=1$ and $q=2$ no transition is observed, for $q=3$ the change from the ferromagnetic to the paramagnetic state has a continuous character and for $q>3$ all the transitions are discontinuous (although there are suggestions of a mixed-order phase transition for $q=5$). In the current work, however, the focus is shifted mainly toward the interplay between overall and concentration. From this point of view, we have, in all cases, a discontinuous transition. On the other hand, the setting examined in \cite{Chmiel2017} is subtly different: the topology is that of a partial duplex, i.e., there are two layers with a partial overlap $r$ -- outside the overlap nodes use a standard $q$-neighbor Ising model dynamics but inside the common part in node's state has to be identical. Additionally, the dynamics allows for the change of the state of the node only if the change is suggested on both levels (the so-called AND rule). The results show that apart from $q=1$ and $q=3$ (continuous phase transition) and $q=2$ (very rich and surprising alteration of different types of transition) for all $q \ge 4$ there is a similar pattern: below a certain critical overlap $r^*$ the system undergoes a discontinuous transition while above it -- a continuous one. This might suggest some similarities to the current work, i.e., a critical value of overlap that influences the behavior of the network, but let us point out that there are at least two substantial differences: (1) in \cite{Chmiel2017} there is no initial polarization of different layers, i.e., one starts simply from a fully ordered system, (2) the relation in question concerns $c$ and $T$, unlike the current work.          

Let us direct final remarks to the ramifications of our work concerning the problem of polarization. Although we are not able to report any real-world data that could be used to validate our model, one can still discuss some possible scenarios arising from the obtained results. It is crucial to understand that in the scope of the terminology used in \cite{Gajewski2022}, we are dealing here with two states: (1) polarization, i.e., when the first and second clusters have (on average) different opinions, and (2) radicalization, i.e., one both clusters have the same opinion (additionally we might consider that state in the interconnected cluster could be referred as to a ``neutral consensus''). The first noteworthy thing is the resemblance between $q=3$ and the infinite range model. In both cases, it is possible to maintain a polarized but weak (in the sense of the level of concentration) state for high $T$. Obviously, there is a dramatic difference in the critical value of temperature that is needed to arrive at this point (of the order of $1$ for $q=3$ and ca. $N$ for the infinite range case). On the other hand, the situation for $q=4$ is dramatically different: no matter what thermal noise is introduced to the system, the polarization level is almost the same just before the system switches to a radicalized state. This brings to the front the role of lobby size: when our group of influence is restricted to just three individuals, we need less than $v=1/4$ overlap to switch from polarization to radicalization, but when the lobby size is 5 times larger, we need to have $v=0.4$ to achieve the same goal. Surprisingly, increasing this number further results in a temporal but noticeable decrease in the needed overlap. 
Although this behavior is evident, we can speculate that its origins are connected to the setting in which it is observed, i.e., a strictly homogeneous system. It is then possible that when examined in a setting where the agents (nodes) are characterized with either heterogeneous degree values or the lobby size is sampled from some distribution $p(q)$ \cite{Radosz2017} -- thus more closely resembling an actual social network -- the phenomenon in question might disappear. It follows that one might, in general, consider examining the conditions upon which the phenomena observed in this paper change their character or even vanish. Another option for the extension of this study is to consider hypergraph structures or higher-order networks that have lately brought a lot of attention \cite{Bianconi2021} also in the context of opinion formation processes, such as majority vote \cite{Noonan2021}, voter \cite{Papanikolaou2022} or threshold \cite{Xu2022} dynamics. In this way, the connection between higher-order settings, where we consider multi-node interactions and group dynamics, might constitute a promising direction for further research in the context of $q$-lobby models of opinion formation.

In summary, the obtained results indicate that lobby size positively affects the level of overlap while the temperature acts in the opposite way. Nonetheless, the analysis presented in the paper pinpoints some striking deviations from these general observations, which can stem from the specifc network structure used in this study.       

\begin{acknowledgments}
This research was funded by POB Research Centre Cybersecurity and Data Science of Warsaw University of Technology, Poland within the Excellence Initiative Program—Research University (ID-UB).
\end{acknowledgments}

\appendix
\section{Solutions for $q=1$ and $q=2$}\label{app:app1}
If $q=1$, we arrive at
\begin{equation}
    -x\left[2z + (1 - z)v\right] = 0,
\end{equation}
where $z=\mathrm{e}^{-2/T}$ with only solution $x=0$, for any value $v$ and $T$.\\
~\\
In case $q=2$ we obtain 
\begin{equation}
x\left[(2v-3)z^2-(1+2v)+4(1-z^2)(v-1)^2x^2\right]=0,
\end{equation}
with three solutions
\begin{align}
x_0 ~ ~ &= ~ 0\nonumber\\
x_{1,2} &= \pm \frac{1}{2(v-1)} \sqrt{\frac{1+2v-(2v-3)z^2}{1-z^2}}.
\end{align}
However, as $x_1 > \frac{1}{2}$ and $x_2 < -\frac{1}{2}$ for any values of $v$ and $T$ the only physical solution is $x=0$.

\section{Linear stability analysis}\label{app:lsa}

In order to follow this procedure, we shall first write out equivalents of Eq. (\ref{eq:c1}) for $c_2$ and $c_0$, i.e.,

\begin{align}
\gamma_2^{+} - \gamma_2^{-} = 0
\label{eq:c2}
\end{align}
and
\begin{align}
\gamma_0^{+} - \gamma_0^{-}  = 0
\label{eq:c0}
\end{align}

Now, in order to linearize our equations, we use l.h.s. of Eqs.~(\ref{eq:c1}), (\ref{eq:c2}) and (\ref{eq:c0}), substitute $c_1$, $c_2$ and $c_0$ as
\begin{align}
c_1 \rightarrow \frac{1}{2} + x_1 + \xi_1\nonumber\\
c_2 \rightarrow \frac{1}{2} + x_2 + \xi_2\nonumber\\
c_0 \rightarrow \frac{1}{2} + \xi_0,
\label{eq:trans}
\end{align}
where $x_1$ and $x_2$ are solutions given in Eq. (\ref{eq:q3sol}), and keep only the linear terms with respect to $\xi_1$, $\xi_2$ and $\xi_0$. Matrix $\mathbf{L}$ presenting coefficients of such a linear expansion has the following general form (owing to the symmetry of the problem, i.e., $c_1$ and $c_2$ are interchangeable in $\gamma_0^{+} - \gamma_0^{-}$, $\gamma_1^{+} - \gamma_1^{-}$ and $\gamma_2^{+} - \gamma_2^{-}$ have the same form, except for swapped $c_1 \rightarrow c_2$, $x_2$ is $-x_1$ etc):

\begin{equation}
    \mathbf{L} =     \begin{blockarray}{cccc}
& {\scriptstyle \xi_0} & {\scriptstyle \xi_1} & {\scriptstyle \xi_2}\\
\begin{block}{c(ccc)}
 {\scriptstyle \gamma_0^{+} - \gamma_0^{-}~~} & s & t & t\\
 {\scriptstyle \gamma_1^{+} - \gamma_1^{-}~~} & u & w & 0\\
 {\scriptstyle \gamma_2^{+} - \gamma_2^{-}~~} & u & 0 & w\\
\end{block}
\end{blockarray}.
\end{equation}
where each column and row is labelled, respectively, with relevant $\xi$ coefficient and $\gamma^{+} - \gamma^{-}$ equation. The eigenvalues of $\mathbf{L}$ are then simply

\begin{equation}
\left\{
\begin{aligned}
\lambda_0 ~ ~ &= w\\
\lambda_{1,2} &= \frac{1}{2}\left( s + w \pm \sqrt{8 u t+(w-s)^2}\right).
\label{eq:lambda}
\end{aligned}\right.
\end{equation}

The problem of extracting the critical value for which $x_1$ loses its stability is then equivalent to finding when the largest eigenvalue $\lambda_{max} = 0$, i.e., $w = 0$ or $2ut - ws=0$. The presented reasoning, based on \cite{lambiotte2007majority} is a general framework, not dependent on the value of $q$ or $T$.

\section{Explicit solution for $q=3$ and $T>0$}\label{app:app3}
The elements of matrix $\mathbf{L}$ for $q=3$ in the $T > 0$ setting read ($z=\mathrm{e}^{-\frac{2}{T}}$ in further formulas)
\begin{widetext}
\begin{equation}   
\left\{
\begin{aligned}
s &= \frac{(v+1)z^3 + 3z + 4-5v}{2(v-2)}\\
t &= \frac{3(v-1)}{4(v-2)}\left(2-z-z^3\right)\\
u &= \frac{3v}{2(v-1)}\frac{(v-1)^2(z^5+z^4)-z^3(v^2-4)-z^2(v^2-8)-4z(v^2-3v+1)+4v(v-1)}{(v-4)(z^2+z)+2(1-v)}\\
w &= \frac{1}{2}\left[ 2(3v-1)+z^3(4-3v)-3z(v-2)\right]
\end{aligned}\right..
\end{equation}
\end{widetext}
Using the relation $2ut-ws=0$ leads to the cubic equation
\begin{equation}
av^3+bv^2+cv+d=0
\label{eq:v3Tcubic}
\end{equation}
that allows to obtain the critical value of $v_3^*(T)$. Coefficients $a$, $b$, $c$ and $d$ depend only on temperature via $z$
\begin{widetext}
\begin{equation}   
\left\{
\begin{aligned}
a &= 3(z^2-1)(z+2)(z^2+z+2)(2z^3+3z-1)\\
b &= (1-u) \left[23 (u+2) z^6+84 z^5+154 z^4+89 z^3+120 z^2-76 z-56\right]\\
c &= 12 \left(17 z^5+5 z^4+7 z^3-22 z^2+3 \left(z^2+z+3\right) z^6-13 z+7\right)\\
d &= -4 (z+1) ((z-1) z+4) \left(2 z^3+3 z-1\right) (2 z (z+1)-1)
\end{aligned}\right.
\end{equation}
\end{widetext}


\section{Concentration expansion in infinite range case}\label{app:app4}
Owing to the fact that we can expand $\ln y$ and $\ln(1-y)$ around $y_0=1/2$ using the Taylor series as
\begin{equation}
    \ln y\bigg\rvert_{y_0=\frac{1}{2}} = -\ln2 - \sum_{n=1}^{n=\infty}(-1)^n \frac{2^n}{n}\left(y-\frac{1}{2}\right)^n
\end{equation}
and
\begin{equation}
    \ln (1-y)\bigg\rvert_{y_0=\frac{1}{2}} = -\ln2 - \sum_{n=1}^{n=\infty}\frac{2^n}{n}\left(y-\frac{1}{2}\right)^n
\end{equation}
we arrive at
\begin{equation}
\ln\frac{y}{1-y}\bigg\rvert_{y_0=\frac{1}{2}} = \sum_{n=1}^{\infty}\frac{2^n\left[1-(-1)^n\right]}{n}\left(y-\frac{1}{2}\right)^n
\label{eq:logtay}
\end{equation}
whose components zeroth for even values of $n$. In particular, we can use Eq.~(\ref{eq:logtay}) use with $c_1 \rightarrow x + 1/2$ transformation, to obtain the following 3rd and 5th order approximations of $\ln\frac{c_1}{1-c_1}$, i.e.,
\begin{equation}
\ln\frac{c_1}{1-c_1} = 2x + \frac{16}{3}x^3 + O(x^5)
\end{equation}
and
\begin{equation}
\ln\frac{c_1}{1-c_1} = 2x + \frac{16}{3}x^3 + \frac{64}{5}x^5 + O(x^7).
\end{equation}
These, in turn, can be used to find solutions of the first equation in (\ref{eq:mf}) with $c_0 \rightarrow 1/2$, i.e.,

\begin{equation}
\frac{2 N}{T}\frac{1-v}{2-v}(2c_1-1) = \ln\frac{c_1}{1-c_1}
\end{equation}
which is simply a slightly rearranged Eq.~(\ref{eq:cmf}). The obtained approximate solutions $c^{(3)}_1$ and $c^{(5)}_1$ read:
\begin{equation}
c^{(3)}_1 = \frac{1}{2}+\frac{\sqrt{3} \sqrt{\frac{N}{T} v-\frac{N}{T}-v+2}}{2 \sqrt{v-2}}
\label{eq:appc3}
\end{equation}
and
\begin{equation}
c^{(5)}_1=\frac{1}{2}+
\frac{\sqrt{6}}{12}\sqrt{\sqrt{\frac{5(36 \frac{N}{T} v-36\frac{N}{T}-31 v+62)}{v-2}}-5}.
\label{eq:appc5}
\end{equation}

Series expansion of $\ln y / \ln(1-y)$ is also the key step to perform linear stability analysis in the case of the mean-field approach. Indeed, if we expand  $\ln c_1 / \ln(1-c_1)$ at $c_1 = 1/2 + x$ and $\ln c_0 / \ln(1-c_0)$ at $c_0 = 1/2$ up to the 5th order, plug the obtained formulas into Eq. (\ref{eq:mf}) and follow the procedure described in Appendix \ref{app:lsa}, we shall arrive at rather concise elements of matrix $\mathbf{L}$   
\begin{equation}
\left\{
\begin{aligned}
s &= \frac{4 \frac{N}{T} v}{2-v}-4\\
t &= \frac{4 \frac{N}{T} (1-v)}{2-v}\\
u &= \frac{4 \frac{N}{T} v}{2-v}\\
w &= \frac{4 \frac{N}{T} (1-v)}{2-v}-\frac{4}{1-4x^2}
\end{aligned}\right.
\end{equation}
At this point, it is straightforward to find the critical value of $v$ by substituting $1/2+x$ with either the numerical solution of (\ref{eq:cmf}) or approximate solutions $c^{(3)}_1$ and $c^{(5)}_1$ and solving $2ut - ws=0$ for $v$.

\bibliography{ref}
\end{document}